\documentclass[fleqn,usenatbib]{mnras}

\usepackage{newtxtext,newtxmath}

\usepackage[T1]{fontenc}

\DeclareRobustCommand{\VAN}[3]{#2}
\let\VANthebibliography\thebibliography
\def\thebibliography{\DeclareRobustCommand{\VAN}[3]{##3}\VANthebibliography}

\usepackage{graphicx}	
\usepackage{amsmath}	
\title[NGC 4839 group]{A deep dive: \textit{Chandra} observations of the NGC 4839 group falling into the Coma cluster}

\author[M. S. Mirakhor et al.]{
M. S. Mirakhor,\thanks{Email: msm0033@uah.edu}
S. A. Walker and 
J. Runge
\\
Department of Physics and Astronomy, The University of Alabama in Huntsville, 301 Sparkman Drive NW, Huntsville, AL 35899, USA
}

\date{Accepted 2023 April 6. Received 2023 March 23; in original form 2022 August 15}

\pubyear{2015}

\begin{document}
\label{firstpage}
\pagerange{\pageref{firstpage}--\pageref{lastpage}}
\maketitle

\begin{abstract}
Cosmological simulations of structure formation predict that galaxy clusters continue to grow and evolve through ongoing mergers with group-scale systems. During these merging events, the ram pressure applied by the intracluster medium acts to strip the gas from the infalling groups, forming large tails of stripped gas, which eventually become part of the main cluster. In this work, we present a detailed analysis of our new deep \textit{Chandra} observations of the NGC 4839 group falling into the nearby Coma cluster, providing a unique opportunity to explore the way galaxy clusters in the local universe continue to grow. Our analysis reveals a cold front feature at the leading head of the group, preceded by a bow shock of hot gas in front with a Mach number of $\sim\! 1.5$. The power spectrum of surface brightness fluctuations in the tail shows that the slope gets less steep as the distance from the leading head increases, changing from $-2.35_{-0.06}^{+0.07}$ at the inner part of the tail to $-1.37_{-0.07}^{+0.09}$ at the outermost part of the tail. These values are shallower than the slope of the Kolmogorov 2D power spectrum, indicating that thermal conduction is being suppressed throughout the tail, enabling long-lived small-scale turbulence, which would typically be washed out if thermal conduction was not inhibited. The characteristic amplitude of surface brightness fluctuations in the tail suggests a mild level of turbulence with a Mach number in the range of 0.1--0.5, agreeing with that found for the infalling group in Abell 2142.
\end{abstract}




\section{Introduction}
Cosmological simulations indicate that present-day galaxy clusters predominantly grow in mass through ongoing mergers with galaxy groups (e.g. \citealt{Dolag09}). Ram-pressure acts to remove the gas from the infalling structures, forming large tails of stripped gas, which ultimately virialize and become part of the cluster. Due to the low X-ray surface brightness of these galaxy groups, detailed studies are challenging and limited to just a handful of systems in which groups are seen to be falling into clusters. These include the Coma cluster (\citealt{Neumann2001}), Abell 85 (\citealt{Durret2005,Ichinohe2015}), Abell 2142 (\citealt{Eckert2014,Eckert2017}), Hydra A (\citealt{DeGrandi2016}), and Abell 1367 \citep{Ge2021}.

Detailed studies of infalling groups provide powerful probes of the way galaxy clusters continue to grow, whilst also revealing the micro-physical properties of the intracluster medium (ICM) in a unique way. By studying the tails of ram-pressure-stripped gas and comparing them to numerical simulations, we can put constraints on the ICM viscosity and magnetic fields that are unattainable by any other means. The velocity difference between the infalling group and the ICM should lead the interface between them to be subject to Kelvin-Helmholtz instabilities (KHIs), the morphology of which depends on the viscosity of the gas. A higher viscosity suppresses the formation of KHIs, creating long X-ray tails \citep{Roediger15,Roediger2015II}. On the other hand, a lower viscosity allows KHIs to quickly form, leading to a significant level of turbulence in the wake behind the infalling group and reducing the overall length of the tail that can form.

The survival of the long tails of infalling groups in the surrounding ICM indicates that thermal conductivity is suppressed. However, one outstanding question is whether thermal conduction is suppressed only at the interface between the infalling group and the ICM, or is suppressed throughout the tail. Recently, \citet{Eckert2017} have analysed \textit{Chandra} observations of an infalling group in the outskirts of the massive cluster Abell 2142. They found that the infalling group possesses an elongated tail that stretches several hundred kiloparsecs and flares out to a width of $\sim\!300$ kpc. By studying the power spectrum of surface brightness fluctuations, \cite{Eckert2017} concluded that thermal conduction is inhibited throughout the tail. Interestingly, this suppression extends beyond the region where magnetic draping should be effective. In addition to plasma micro-instabilities, it is argued that the cascading turbulent motions that cause a small-scale tangled magnetic field configuration may be responsible for this suppression (\citealt{Rechester78,Chandran98,Rusz10,Kom14}).


\begin{figure*}
    \centering
    \includegraphics[width=1.0\textwidth]{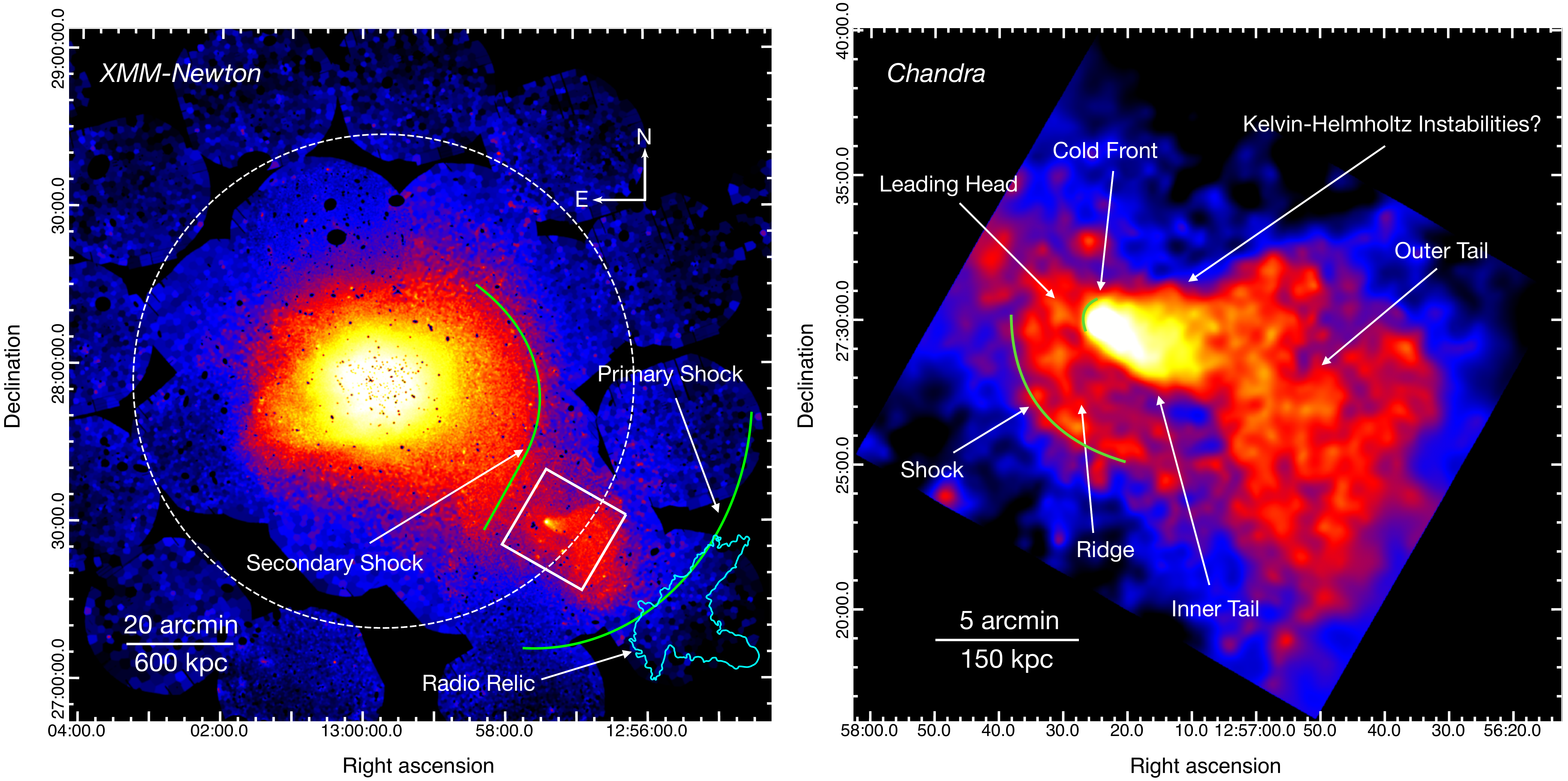}
    \vspace{-4mm}
    \caption{Left: Voronoi tessellated, \textit{XMM--Newton} mosaicked image of the Coma cluster in the 0.7--1.2 keV energy band (taken from \citealt{Mira20}), showing the infalling group NGC 4839 to the southwest of the cluster. The white box highlights the spatial extent of our \textit{Chandra} observation of the group. The white dashed circle marks the radius of $r_{500}$ from the cluster centre. The green curves are the presumed positions of the primary and secondary shocks (taken from Fig. 11 in \citealt{Churazov2021}), and the cyan contour marks the location of the radio relic (taken from \citealt{Bonafede2022}). These features are discussed in Section~\ref{sec: dynamics}. Right: Exposure-corrected, background-subtracted \textit{Chandra} image of the infalling group NGC 4839, smoothed heavily to highlight the diffuse emission from the group. The  morphological features discussed in the main text are labelled in the image. The green curves mark the positions of cold and shock fronts detected in our \textit{Chandra} data.} 
    \label{fig:flux_image}
\end{figure*}

The infalling group NGC 4839 in the outskirts of the nearby Coma cluster is one of the lowest-redshift examples of a group falling into a massive cluster with a striped tail. It lies at a distance of around 40 arcmin (corresponding roughly to 1.1 Mpc) to the southwest of the Coma centre. The \textit{XMM{--}Newton} image of the NGC 4839 group (Fig. \ref{fig:flux_image}, left) by \citet{Mira20} reveals a very long tail of ram-pressure-stripped gas extending for at least 300 kpc, preceded by an X-ray bright head-like structure. The \textit{XMM--Newton} image also reveals the group falling into the Coma core with its tail pointing away from the cluster, coinciding with where a cosmic web filament joins the Coma cluster to Abell 1367 \citep{Malavasi2020}.

It has long been debated whether the NGC 4839 group is falling into Coma for the first time \citep[e.g.][]{Neumann2001,Akamatsu2013} or it has already passed the Coma centre and is now on its second infall \citep[e.g.][]{Burns1994,Oh2023}. \citet{Lyskova2019} have compared the observed X-ray geometry of the NGC 4839 group with a series of numerical simulations of different infall scenarios, and have found the best match to be a scenario in which the NGC 4839 group is on its second infall. In this case, the group would have entered the Coma cluster from the northeast, before passing close by the core and heading out to the southwest, where it has turned around and begun to fall back towards the core. These simulations indicate that the group has spent an enormous amount of time in the ICM of the cluster ($\sim\! 2.5$ Gyr), yet its tail has still survived against thermal conduction.

In this paper, we examine several X-ray properties of the infalling group NGC 4839 and probe the suppression of thermal conduction throughout its tail using our new deep \textit{Chandra} observations. The NGC 4839 group shares a similar morphology to that of Abell 2142, providing a rare opportunity to explore the way galaxy clusters in the local universe continue to grow. Furthermore, the large-spatial extent of the NGC 4839 group, afforded by the low redshift of the system, allows us to explore the microphysics of the merging process on unprecedentedly small scales. The rest of the paper is structured as follows: Section~\ref{sec: analysis procedure} goes over the observations, imaging, and spectral analyses; Sections~\ref{sec: head and ridge} and~\ref{sec: khi} analysis the main features observed in our \textit{Chandra} data; Section~\ref{sec: tail} examines the power spectrum of surface brightness fluctuations in the tail to probes turbulence and transport processes;  our findings are discussed in Section~\ref{sec: discussion}; and our conclusions are presented in Section~\ref{sec: conclusions}.

Throughout the paper, for consistency with \citet{Eckert2017}, we adopt a $\Lambda$CDM cosmology with $\Omega_{\rm{\Lambda}}=0.7$, $\Omega_{\rm{m}}=0.3$, and $H_0=70$~km s$^{-1}$ Mpc$^{-1}$. At the redshift of NGC 4839 \citep[$z = 0.0245$;][]{Ahn2012}, 1 arcsec corresponds to 0.496 kpc. Uncertainties are at the 68 per cent confidence level unless otherwise stated.

\section{Analysis procedure}
\label{sec: analysis procedure}

\subsection{Imaging}
\label{sec: data}
We used our new seven \textit{Chandra} observations taken in 2020 (PI: S. A. Walker) and one archival observation from 2010 for a total observation time of 232 ks. Table~\ref{table:observations} gives a summary of the \textit{Chandra} observations used in this work. All data were reprocessed using the \texttt{chandra\_repro} script of \texttt{CIAO 4.13} and \texttt{CALDB 4.9.1}. Background light curves were analyzed to filter out any time periods affected by flares. Counts, exposure maps, and flux images were produced from the combined observations using the \texttt{CIAO} script \texttt{merge\_obs}. Point sources were detected and excluded from the analysis using \texttt{CIAO wavdetect} with wavelet scales of 1, 2, 4, 8, and 16 pixels.

Following the same procedure as described in \cite{HM06}, we used the stowed ACIS images with a total observation time of 367 ks in order to model and subtract instrumentation contribution to the background. The stowed images were scaled so that the 9.5--12 keV count rates matched the observations. 

The right-hand panel of Fig. \ref{fig:flux_image} shows our exposure-corrected, background-subtracted \textit{Chandra} image of the NGC 4839 group in the 0.5--7.0 keV band, smoothed heavily to highlight the diffuse emission from the group. An incredible variety of structures are revealed by our deep \textit{Chandra} observations. It shows an elongated tail of ram-pressure-stripped gas pointing away from the Coma centre, preceded by an X-ray bright structure, the head. The extended tail can be divided into two main regions: inner and outer. The northern side of the inner tail exhibits a ripple feature around 25 kpc wide that resembles a KHI feature (this is labelled on the right-hand panel of Fig. \ref{fig:flux_image} as `Kelvin-Helmholtz Instabilities?'). We discuss this feature further in Section \ref{sec: khi}. There is also a curved ridge of enhanced X-ray surface brightness in front and to the southeast of the head, which appears separated from it by a dark trough. In \citet{Lyskova2019}, this emission was referred to as the "sheath" and is explained by the interaction between the re-infalling group and the gas stripped from its tail, which is now mixed with the surrounding ICM.

\begin{table}
     \centering
     \caption{Summary of the \textit{Chandra} observations for the NGC 4839 group.}
        \begin{tabular}{ccccc}
     \hline
     Obs. ID & Date & RA  & Dec.  & Exposure (ks) \\
      \hline
      \vspace{-1.5 em} \\
        22648  & 2020-03-03  & 12 57 11.32 & +27 27 40.14 & 33 \\
        22649  & 2020-03-04  & 12 57 11.32 & +27 27 40.14 & 34 \\
        22930  & 2020-08-11  & 12 57 00.26 & +27 28 00.59 &	37 \\
        23182  & 2020-03-04  & 12 57 11.32 & +27 27 40.14 & 31 \\
        23361  & 2020-11-02  & 12 57 00.26 & +27 28 00.59 & 19 \\
        24853  & 2020-11-03  & 12 57 00.26 & +27 28 00.59 & 27 \\
        24854  & 2020-11-08  & 12 57 00.26 & +27 28 00.59 & 8  \\
        12887  & 2010-11-11  & 12 57 24.10 & +27 29 47.00 & 43 \\
        \hline
      \end{tabular}      
    \label{table:observations}
\end{table}

\subsection{Spectral analysis}
\label{sec: spectral analysis}
To study the spectral properties of the gas in various regions of the NGC 4839 group, it is necessary to choose spatial regions for X-ray spectroscopy. To do that, we used the contour binning software \texttt{CONTBIN} version 1.6 \citep{Sanders06} with a threshold for the signal-to-noise ratio of 50. This signal-to-noise ratio was used to accommodate the fainter parts of the field. The software chooses regions using contours from adaptively-smoothed, background-subtracted images in such a way that the generated regions closely follow the X-ray surface brightness distribution. For each selected region, we run the \texttt{CIAO} tool \texttt{specextract} to create the X-ray spectrum for the source along with the response matrix files (RMFs) and the ancillary response files (ARFs). 

For the spectral fitting and background subtraction we followed the methods discussed in \citet{Wang2014} and \citet{Walker2018NatAs}. First, stowed backgrounds\footnote{\url{https://cxc.harvard.edu/contrib/maxim/acisbg/}} were used, scaled to match the 9.5--12 keV count rate, to subtract the particle background. To model the soft X-ray background, we used \textit{ROSAT} All Sky Survey data for a background annulus around the Coma cluster extending from 2--3 degrees. The soft background was modelled as consisting of an absorbed \texttt{APEC} model originating from the Galactic Halo with a temperature of $0.23\pm0.01$ keV and normalisation of $2.7_{-0.3}^{+0.3}\times 10^{-3}$, and an unabsorbed \texttt{APEC} component from the local hot bubble with a temperature of $0.11 \pm 0.001$ keV and normalisation of $1.7_{-0.4}^{+0.4}\times 10^{-3}$ (the quoted normalisations are in units of $\frac{10^{-14}}{4\pi[D_{\rm{A}}(1+z)]^2}$ cm$^{-5}$ per 400$\pi$ square arcmin area, where $D_{\rm{A}}$ is the angular diameter distance). These values are consistent with those found in \citet{Simi13}. For the cosmic X-ray background component from unresolved point sources, this was modelled as an absorbed power law of index 1.4 and allowed to be a free parameter in the spectral fitting. After modelling the background, the emission from the galaxy group was fitted with an absorbed \texttt{APEC} model. The Galactic column density was fixed at  $N_{\textup{H}}=8.79\times10^{19}$ cm$^{-2}$ (\citealt{DL90}), the metallicity fixed at $Z = 0.3 Z_\odot$ (as was done in \citealt{Simi13}), and the redshift fixed at $z = 0.0245$.  

\begin{figure}
    \centering
    \includegraphics[width=\columnwidth]{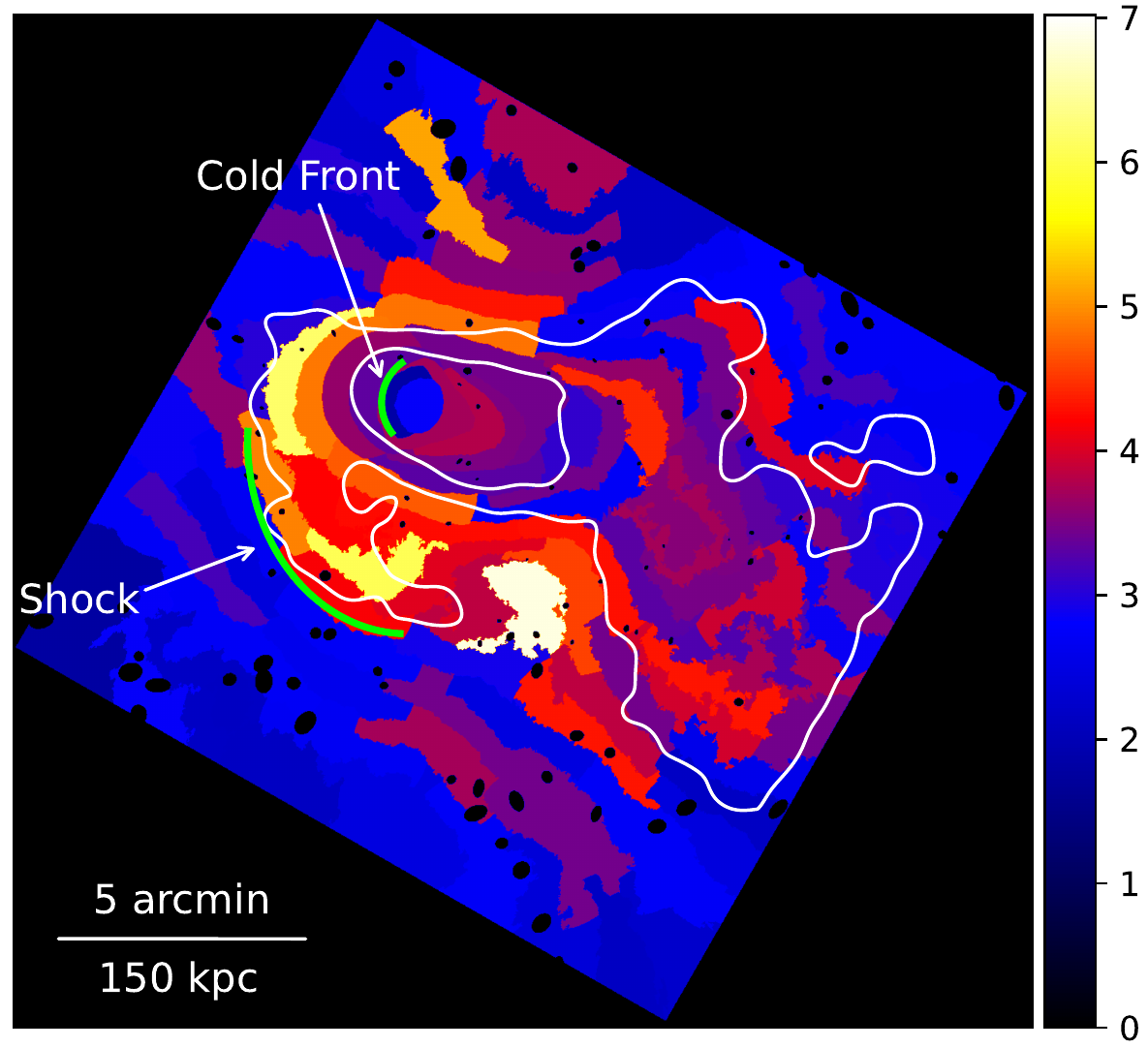}
    \vspace{-3mm}
    \caption{Temperature map of the infalling group NGC 4839 created using the contour binning technique with a signal-to-noise ratio of 50. The green curves mark the positions of cold and shock fronts discussed in the main text. White contours are the X-ray emission shown in Fig.~\ref{fig:flux_image}.}   
    \label{fig:temp_map}
\end{figure}

The temperature map of the NGC 4839 group is shown in Fig. \ref{fig:temp_map}. Some interesting features can be seen in the temperature map. A region of cold gas ($\sim \!2$ keV) can be seen within the leading head, preceded by what appears to be a region of shock-heated gas ($>\! 5$ keV) in front and to the southeast. Interestingly, this hot gas extends along the southern side of the outer tail. The gas temperature in the outer tail is $\sim\!3.5$ keV, agreeing with that obtained using \textit{Suzaku} \citep{Akamatsu2013}. In the following sections, we discuss in detail the properties of the features reported here.

\begin{figure}
    \centering
    \includegraphics[width=\columnwidth]{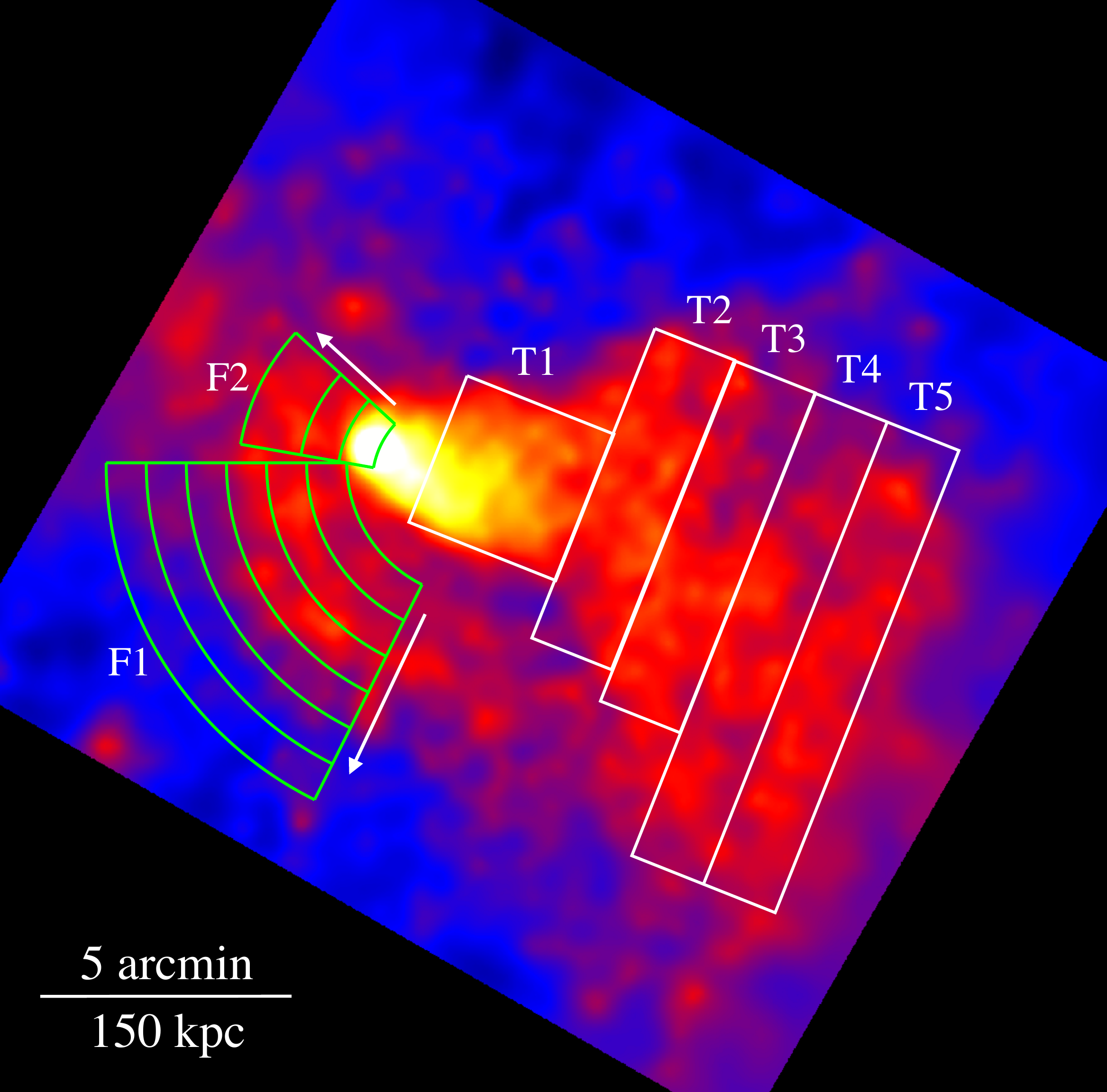}
    \vspace{-3mm}
    \caption{Exposure-corrected \textit{Chandra} X-ray image of the NGC 4839 group, showing the locations of regions used to analyse its properties. The green annuli mark the regions where the properties of the head and ridge features were studied, while the white boxes mark the regions where the power spectrum of surface brightness fluctuations in the tail was extracted. The arrows indicate the direction in which the surface brightness and temperature profiles were extracted. } 
    \label{fig:tailregion}
\end{figure}

\section{The head and ridge features}
\label{sec: head and ridge}

As shown in Fig. \ref{fig:flux_image} (right) and Fig. \ref{fig:temp_map}, there are sharp jumps in the surface brightness and the gas temperature in front and to the southeast of the head structure, suggesting the presence of a leading edge. To investigate this possibility further, we extracted the surface brightness and temperature profiles from a region shown in Fig. \ref{fig:tailregion} (labelled F1). The extracted surface brightness profile was then fitted with a broken power-law model projected along the line of sight \citep{MV07} to determine a possible density discontinuity across the selected region. Assuming spherical symmetry, the density distribution can be given by
 \begin{align}
    n(r) = \Biggl\{
        \begin{array}{ll}
        n_0\Bigl(\frac{r}{r_{\rm{sh}}}\Bigl)^{\alpha_1} & \text{if } r \leq r_{\rm{sh}}, \\
        \frac{1}{C}n_0\Bigl(\frac{r}{r_{\rm{sh}}}\Bigl)^{\alpha_2} & \text{if } r > r_{\rm{sh}},
        \end{array}
    \label{eq: bkn_pow_law}
\end{align}
where $n_0$ is the density normalisation, $\alpha_1$ and $\alpha_2$ are the power-law indices, and $r_{\rm{sh}}$ is the shock putative distance where the surface brightness discontinuity is located. At the discontinuity location, the post-shock density, $n_2$, is higher than the pre-shock density, $n_1$, by a factor of $C=n_2/n_1$. This factor is related to the Mach number through the Rankine-Hugoniot jump conditions \citep{landau1987}.  The Mach number can be calculated either using the density discontinuity:  
\begin{equation}
\mathcal{M} = \left [\frac{2C}{\gamma + 1- C \left ( \gamma -1 \right ) }  \right ]^{\frac{1}{2}},
\label{eq: density_jump}
\end{equation}
or temperature discontinuity:
\begin{equation}
\mathcal{M} = \left [ \frac{(\gamma +1)^2(\frac{T_2}{T_1}-1)}{2\gamma(\gamma-1)} \right ] ^{\frac{1}{2}},  
\label{eq: temp_jump}
\end{equation}
where $T_1$ and $T_2$ are, respectively, the pre-shock and post-shock temperatures at the location of the temperature discontinuity, and $\gamma=5/3$ for a non-relativistic monatomic gas.

\begin{figure}
    \centering
    \includegraphics[width=.83\columnwidth]{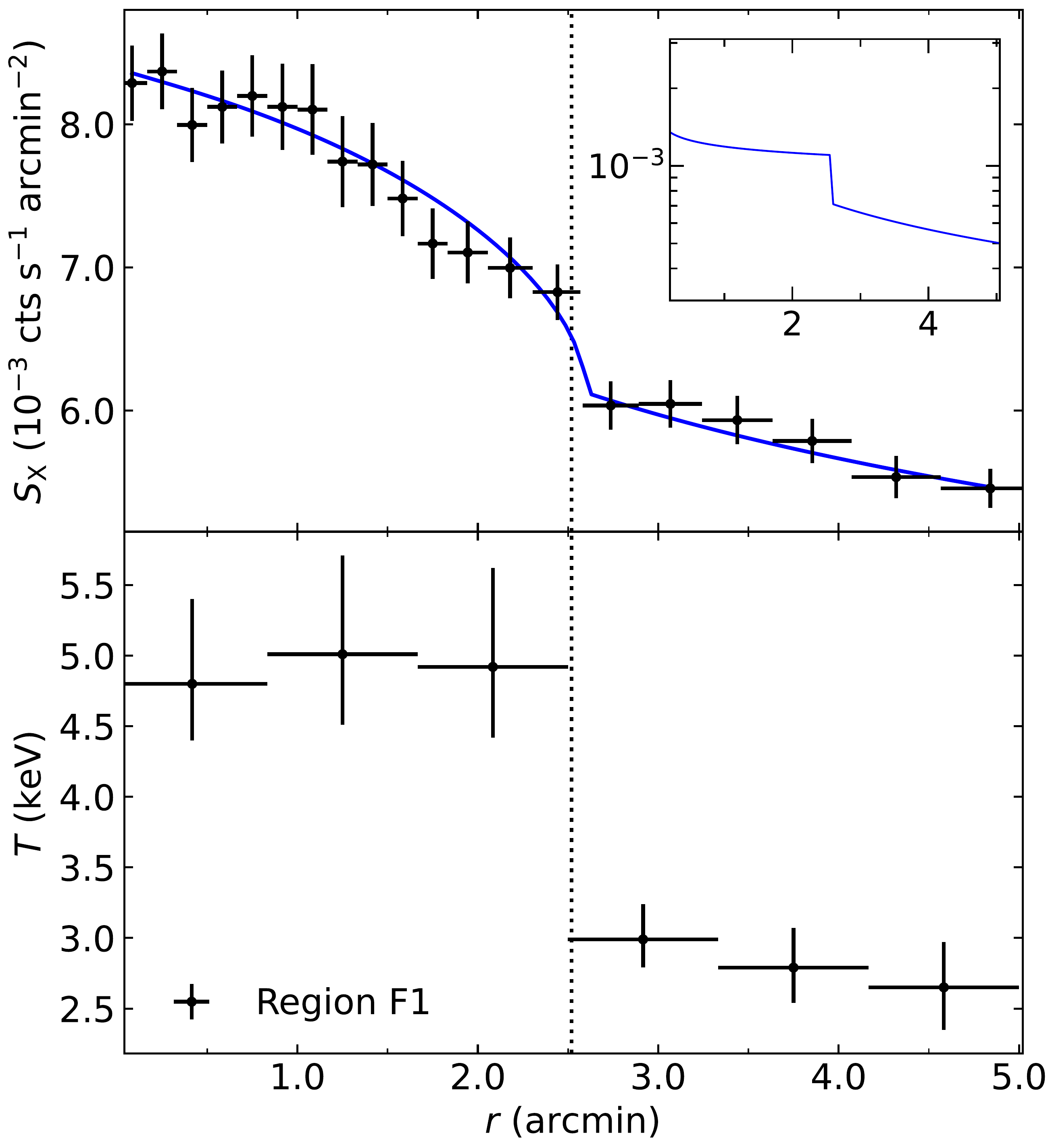}
    \caption{Top: Surface brightness profile extracted from the green annuli labelled F1 in Fig.~\ref{fig:tailregion}. The best fit to the data using a broken power-law model is shown as a solid blue line. The inset shows the gas density profile, where its x-axis is in units of arcmin and the y-axis is in unis of cm$^{-3}$. Bottom: Temperature profile for the same region as noted above. The dotted vertical line shows the location of the shock edge.  } 
    \label{fig:SB_temp_prof}
    \vspace{-4mm}
\end{figure}
\begin{figure}
    \centering
    \includegraphics[width=.83\columnwidth]{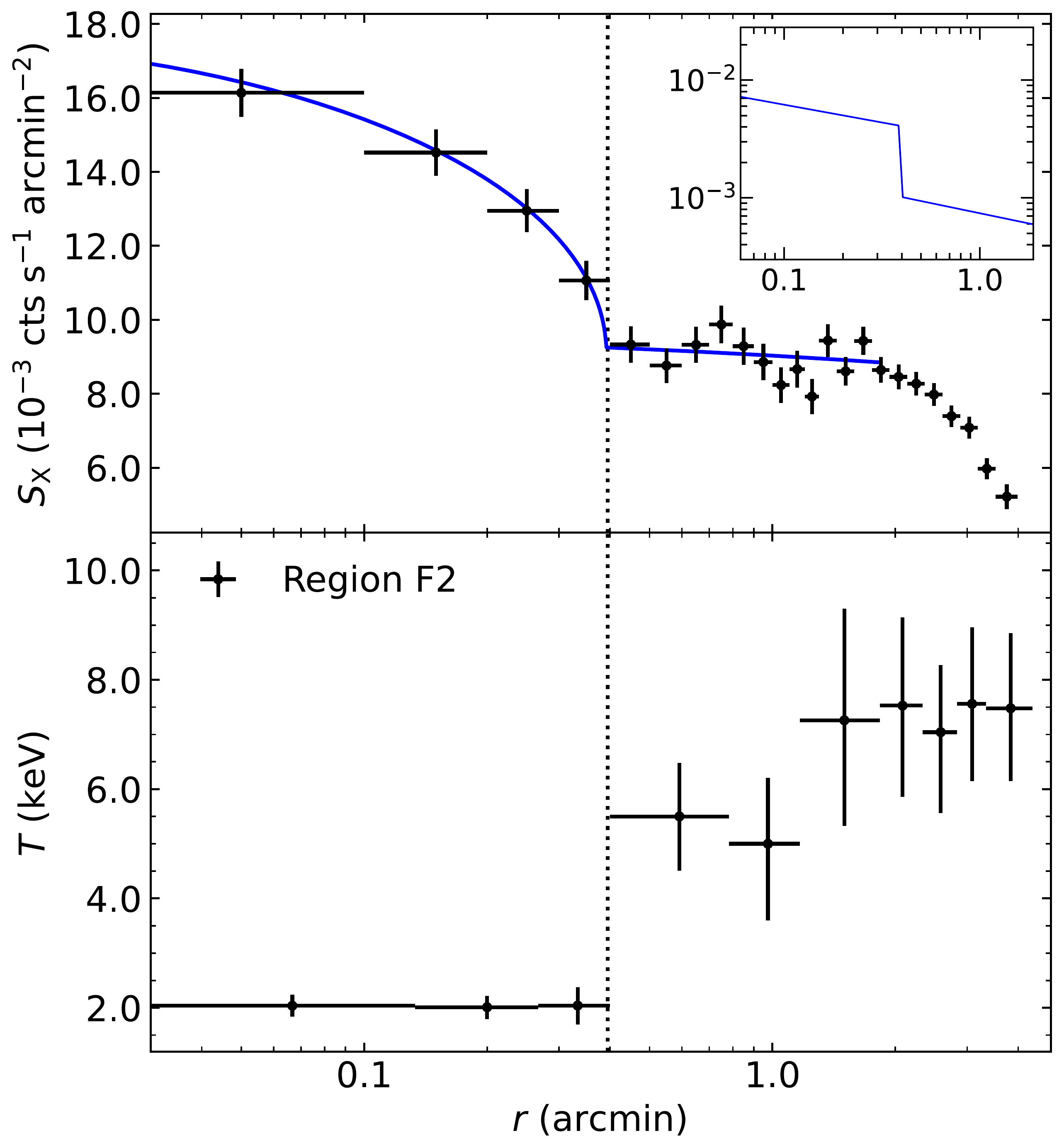}
    \caption{Top: Surface brightness profile extracted from the green annuli labelled F2 in Fig.~\ref{fig:tailregion}. The best fit to the data using a broken power-law model is shown as a solid blue line. The inset shows the gas density profile, where its x-axis is in units of arcmin and the y-axis is in unis of cm$^{-3}$. The surface brightness profile drops off beyond 2 arcmin, indicating the flat plateau outside the cold front is not the background level.  Bottom: Temperature profile for the same region as noted above. The dotted vertical line show the location of the cold front.    } 
    \label{fig:front_temp_prof}
    \vspace{-3mm}
\end{figure}

\begin{table*}
\begin{minipage}{120mm}
    \centering
     \caption{Best-fitting parameters of the broken power-law model.}
        \begin{tabular}{c c|c|c|c|c|c}
     \hline
     Region & $n_0$ ($10^{-3}$ cm$^{-3}$) & $r_{\rm{sh}}$ (arcmin)  & $\alpha_1$   & $\alpha_2$ & $C$ & Reduced $\chi^2$ \\
      \hline
      \vspace{-1.5 em} \\
      F1 & $1.11_{-0.07}^{+0.09}$ & $2.59_{-0.17}^{+0.18}$ & $-0.09_{-0.01}^{+0.01}$ &  $-0.52_{-0.04}^{+0.05}$ & $1.60_{-0.26}^{+0.28}$ & 1.05 \\
      F2 & $4.03_{-0.31}^{+0.35}$ & $0.39_{-0.04}^{+0.04}$ & $-0.30_{-0.02}^{+0.03}$ &  $-0.35_{-0.04}^{+0.05}$ & $4.07_{-0.97}^{+1.36}$ & 1.08 \\    
        \hline
      \end{tabular}      
    \label{table:brknpow}
\end{minipage}    
\end{table*}

The surface brightness profile and the best-fitting model are shown in the upper panel of Fig.~\ref{fig:SB_temp_prof}. In Table \ref{table:brknpow}, the best-fitting parameters of the broken power-law model are presented. There is a clear discontinuity in the surface brightness profile at 2.6 arcmin. Our best-fitting model indicates a density drop by a factor of $C=1.6$, corresponding to a Mach number of $\mathcal{M}=1.4_{-0.2}^{+0.3}$. At this surface brightness discontinuity, the gas temperature drops by a factor of $T_2/T_1=1.7$ (see Fig.~\ref{fig:SB_temp_prof}, lower panel), resulting in a Mach number of $\mathcal{M}=1.5_{-0.5}^{+0.8}$, in agreement with what we found using the gas density.

We also note a sharp drop in the surface brightness in front of the head structure (Fig. \ref{fig:flux_image}, right). To investigate the nature of this edge, we extracted the surface brightness and temperature profiles from the region labelled F2 in Fig.~\ref{fig:tailregion}. Fitting the surface brightness profile of the region with the broken power-law model (equation \ref{eq: bkn_pow_law}), we found a very sharp drop in the gas density by a factor of $4.1_{-1.0}^{+1.4}$ at a distance of $\sim 0.4$ arcmin (Fig. \ref{fig:front_temp_prof}, upper panel). The surface brightness profile along this direction drops off at large radii, indicating the flat plateau outside the cold front is not the background level. The best-fitting parameters of this fit are shown in Table \ref{table:brknpow}. The location of this sharp density edge coincides with the location of the temperature jump within the head structure (Fig. \ref{fig:front_temp_prof}, lower panel). Given that the gas temperature within the head is significantly lower than its surrounding, we conclude that this feature is a cold front. At this cold front, the gas temperature jumps by a factor of $2.7$. This sharp edge is similar to that reported at the leading edge of the bullet in the Bullet cluster \citep{Markevitch2002}. \citet{Eckert2017} also found a cold front feature associated with a sharp density discontinuity of $4.5_{-1.0}^{+1.6}$ at the leading edge of the infalling group in Abell 2142.

\section{KHI-like feature}
\label{sec: khi}
Cosmological simulations \citep[e.g.][]{Roediger15,Roediger2015II} showed that KHIs can form to the sides of infalling groups in clusters of galaxies. The simulations also indicate that these KHIs introduce surface brightness fluctuations and horn-like features in the ICM. The presence of a sufficiently viscous or magnetised plasma suppresses these KHIs, and hence turbulent mixing. Consequently, under these dynamical conditions, one expects to see an unmixed, bright tail on scales of tens or hundreds of kiloparsecs in the wake behind the infalling group. On the other hand, for an inviscid or weekly magnetised plasma, the stripped gas in the wake is expected to mix quickly with the surrounding ICM via KHIs, suppressing the discontinuity in the gas density and the surface brightness.

\begin{figure}
    \centering
    \includegraphics[width=0.9\columnwidth]{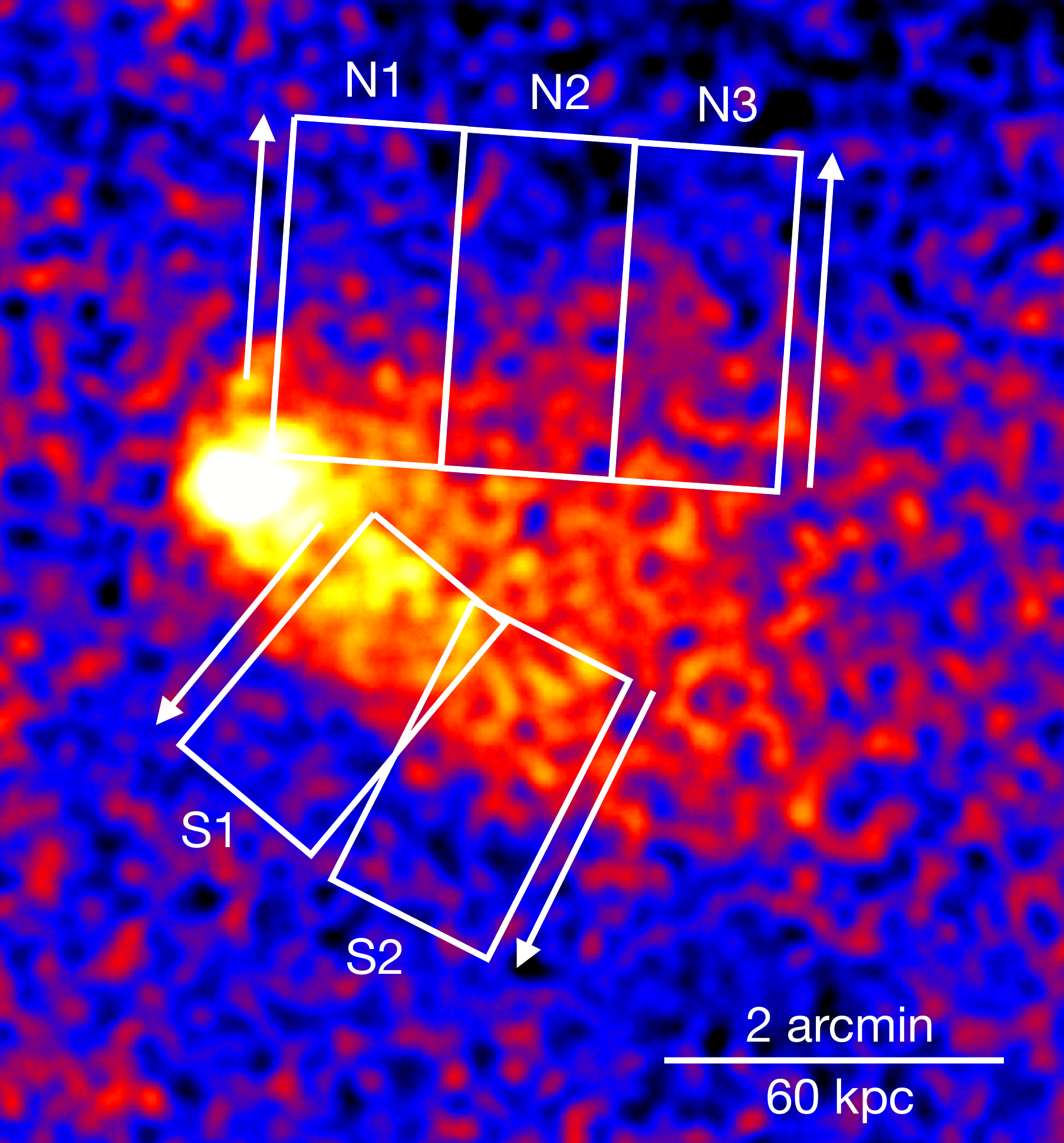}
    \vspace{-1mm}
    \caption{Zoomed-in region of the \textit{Chandra} image of the NGC 4839 group, showing the locations of regions used to probe for surface brightness fluctuations in the ICM. The white arrows indicate the direction in which the surface brightness profiles were extracted.} 
    \label{fig:sb_sides_label}
    \vspace{-4mm}
\end{figure}

\begin{figure}
    \centering
    \includegraphics[width=\columnwidth]{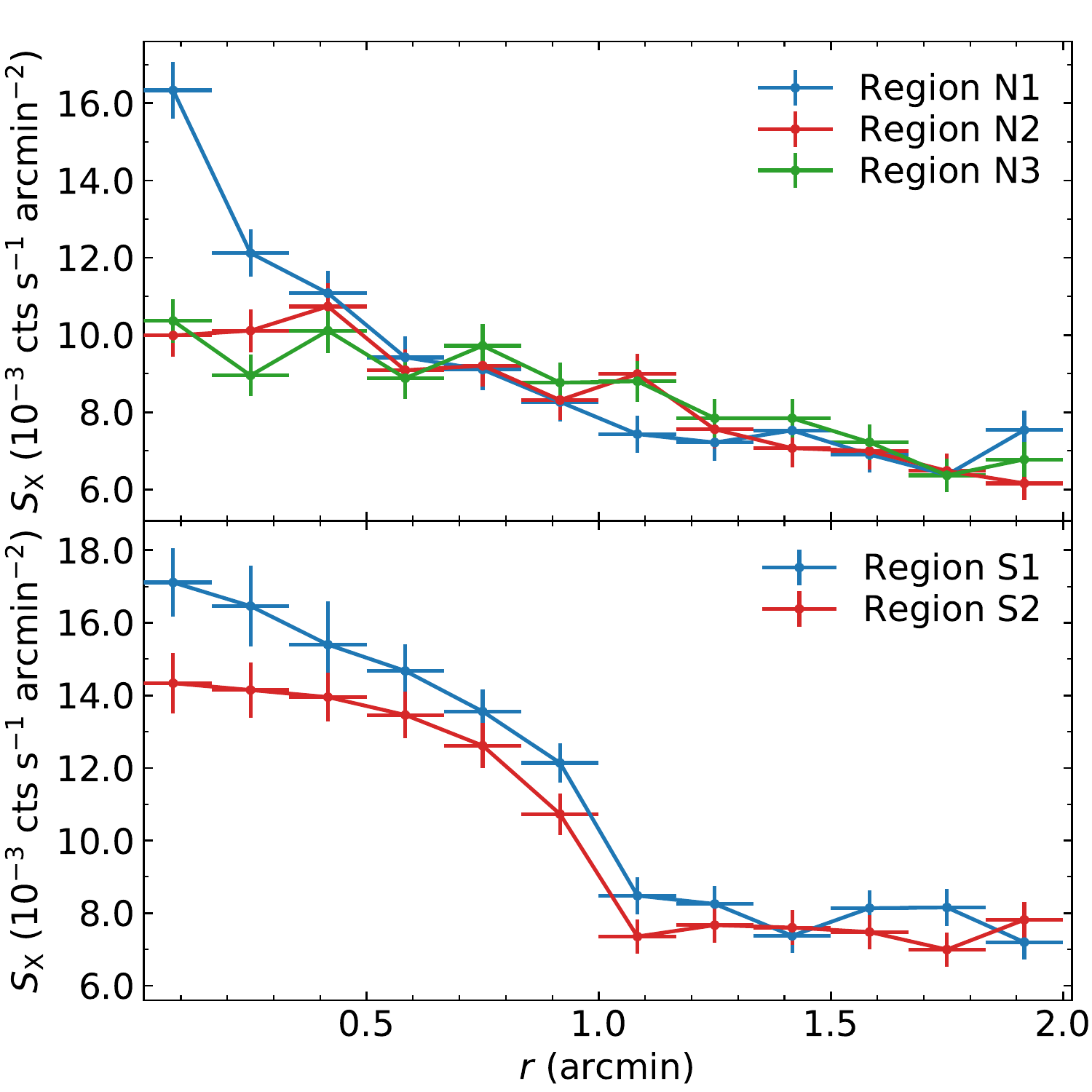}
    \vspace{-6mm}
    \caption{Surface brightness profiles for the northern and southern sides of the infalling group NGC 4839, following the directions of the white arrows shown in Fig.~\ref{fig:sb_sides_label}. While the southern regions show a clear edge as indicated by a steep drop in the surface brightness profiles, the northern regions show no clear edge. } 
    \label{fig:sb_sides}
\end{figure}

\begin{figure*}
    \centering
    \includegraphics[width=\textwidth]{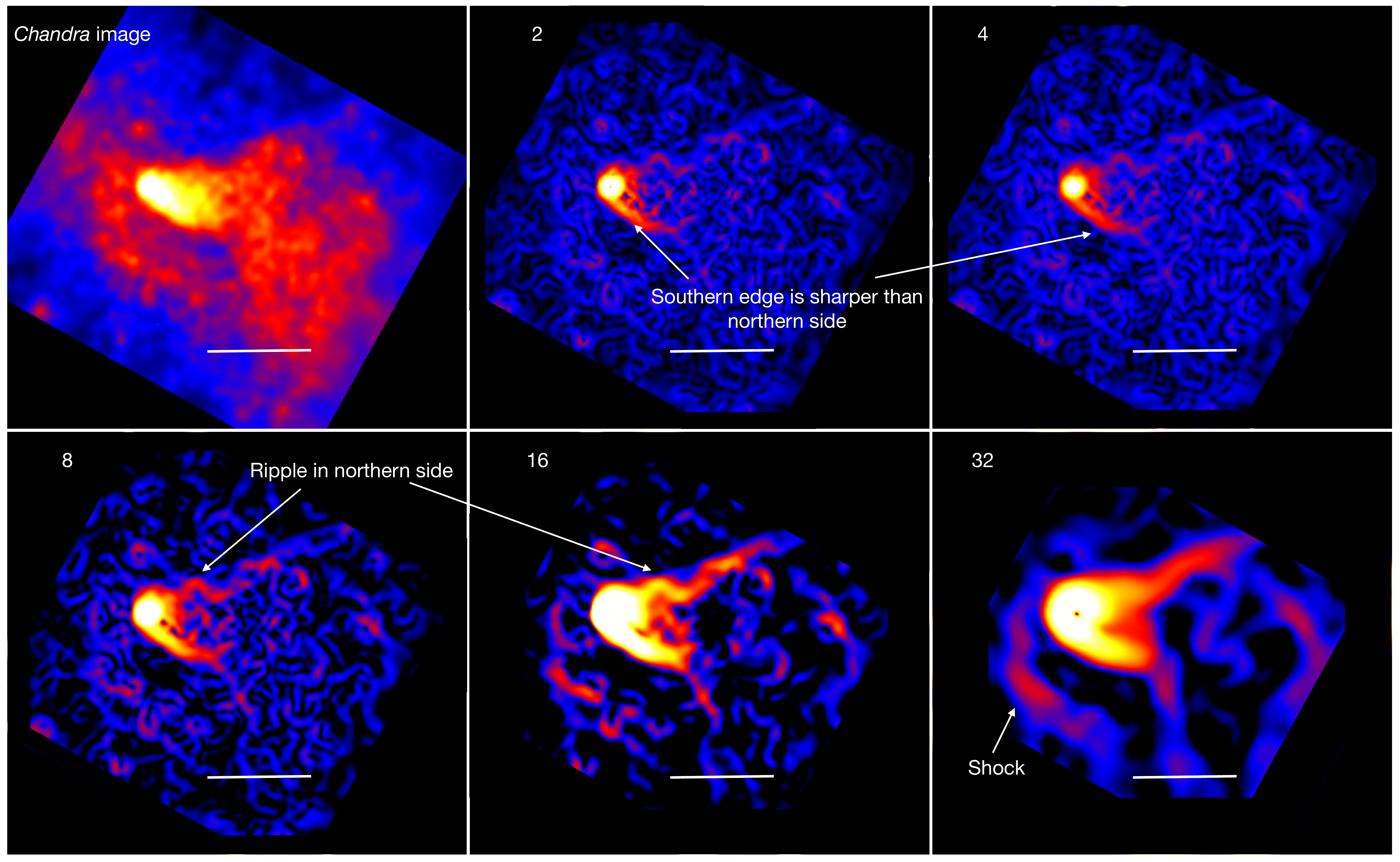}
    \vspace{-4mm}
    \caption{\textit{Chandra} image of the NGC 4839 group along with the GGM filtered images with width scales of 2, 4, 8, 16, 32 pixels. The white bar has a length of 5 arcmin (corresponding to 150 kpc).  }  
    \label{fig:GGM_sides}
\end{figure*}

As shown in our \textit{Chandra} image (Fig.~\ref{fig:flux_image}, right-hand panel), the NGC 4839 group has a straight inner tail along its southern side and a KHI-like feature along its northern side. If this is the case, this feature should appear in the surface brightness profile. For this purpose, we extracted surface brightness profiles from the northern and southern sides of the inner tail. Fig.~\ref{fig:sb_sides_label} shows the regions in which surface brightness profiles were extracted: three in the north and two in the south. The surface brightness profiles are presented in Fig.~\ref{fig:sb_sides}. While the southern regions show an edge as indicated by a steep drop in the surface brightness profiles, the northern regions show no clear edge. We also applied the Gaussian Gradient Magnitude (GGM) filter \citep{Sanders2016,Walker2016} to the \textit{Chandra} data of the group to enhance surface brightness edges in the X-ray image. Fig.~\ref{fig:GGM_sides} shows the \textit{Chandra} image of the group along with the GGM-filtered images using five different width scales. In the filtered images with width scales of 2 and 4 pixels, an edge can be clearly seen along the southern side of the inner tail, while a very faint ripple in surface brightness, which coincides with the location of the KHI-like feature, is visible along the northern side of the inner tail. In the image filtered with a scale of 8 pixels, the ripple along the northern side of the inner tail becomes more visible, but its gradient is not as sharp as that along the southern side. Using scales of 16 and 32 pixels, the filtered images become more sensitive to structures at larger radii, including the shock front discussed in Section~\ref{sec: head and ridge}.   



The above findings might suggest that the stripped gas along the northern side of the inner tail is mixed with the surrounding ICM via KHIs, which implies an inviscid or weekly magnetised plasma. On the other hand, the X-ray bright tail that extends for several hundred kiloparsecs in the wake behind the group argues against the presence of KHIs. This apparently unmixed tail might indicate significant viscosity or draped magnetic fields. Alternatively, this structure could be due to the complex dynamic history of the NGC 4839 group. Since the group is heading back to its second infall as suggested by several recent studies \citep[e.g.][]{Lyskova2019}, this peculiar geometry of the tail could be the result of a turbulent motion as the group is moving through its own gas. However, given that this KHI-like feature has a small-spatial extent with a length of $\sim 25$ kpc, the existing data do not allow us to conclusively constrain its properties and thus its nature, as the smallest spatial scales on which we can map the temperature with the current depth of \textit{Chandra} data is around $50\times50$ kpc.

\section{The tail}
\label{sec: tail}

\subsection{Morphology}
\label{sec: morphology}
A diversity of structures can be seen in the tail of the group from our deep \textit{Chandra} observations (Fig. \ref{fig:flux_image}, right-hand panel). The inner tail, the region closest to the head, is relatively straight and extends up to $\sim\! 150$ kpc from the head. However, the northern side of the inner tail does not appear to be as straight as the southern side and exhibits some ripples in the X-ray surface brightness that resemble KHI rolls. Beyond 150 kpc from the head, the inner tail flares outwards into a much larger, turbulent outer tail, which extends over several hundred kiloparsecs. The outermost regions of the tail display an irregular and patchy morphology, and the gas that is associated with the tail is likely to be mixed with the surrounding ICM.


\subsection{Power spectrum}
\label{sec:power spectrum}
Previous studies \citep[e.g.][]{Churazov2012,Gaspari13,Zhuravleva2014,Gaspari14,Hofmann2016} showed that the characteristic amplitude of the power spectrum of X-ray surface brightness fluctuations in the ICM is linearly proportional to the level of turbulent motion. The slope of the power spectrum, on the other hand, is sensitive to the level of diffusion in the ICM, dominated by conduction \citep{Gaspari13}. Therefore, by measuring the power spectrum of surface brightness fluctuations, we can trace the effects of turbulence and transport processes in the ICM. Due to the relatively flat surface brightness distribution and the remarkable level of dynamical activity, the tail of the NGC 4839 group is an ideal target for investigating the X-ray surface brightness fluctuations to highlight the underlying physical properties of the gas associated with the tail. Furthermore, \citet{Eckert2017} have shown that the shape of the power spectrum is essentially unaffected by projection effects when the motion is mostly in the plane of the sky, which is the case for the NGC 4839 group \citep{CD96}.

To extract the power spectrum of the X-ray surface brightness fluctuations, we defined five regions (shown in Fig.~\ref{fig:tailregion}) along the tail of the NGC 4839 group. The goal is to determine how the power spectrum changes with increasing distance from the head of the group. For this purpose, region T1 is set to trace the gas of the inner tail, while the other regions are set to trace the gas of the outer tail. Below, we provide a detailed description of the analytical procedures that were followed to calculate the power spectrum of the surface brightness fluctuations in the tail.


\subsubsection{Method}
\label{sec: method}
Following the method described in \citet{Churazov2012}, \citet{Gaspari13}, and \citet{Eckert2017}, the power spectrum of surface brightness fluctuations was extracted using the modified $\Delta$-variance method introduced by \cite{Arevalo2012}. The raw image ($I$) was smoothed by two Gaussian filters with widths that differ by a small amount. To correct for gaps in the data, the smoothed images were divided by the smoothed exposure map ($E$). The difference between these two images resulted in a convolved image that is dominated by fluctuations at scales of $\sim\!\sigma$, 


\begin{figure*}
    \centering
    \includegraphics[width=\textwidth]{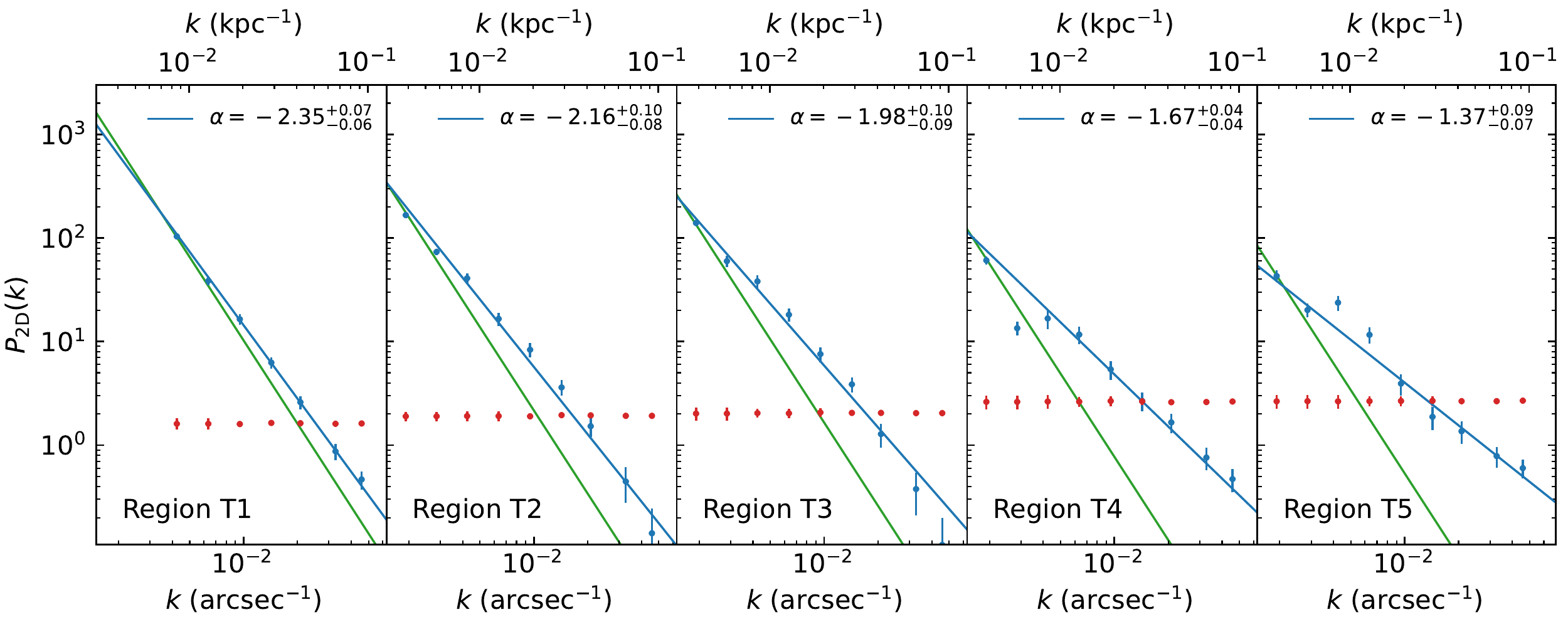}
    \includegraphics[width=\textwidth]{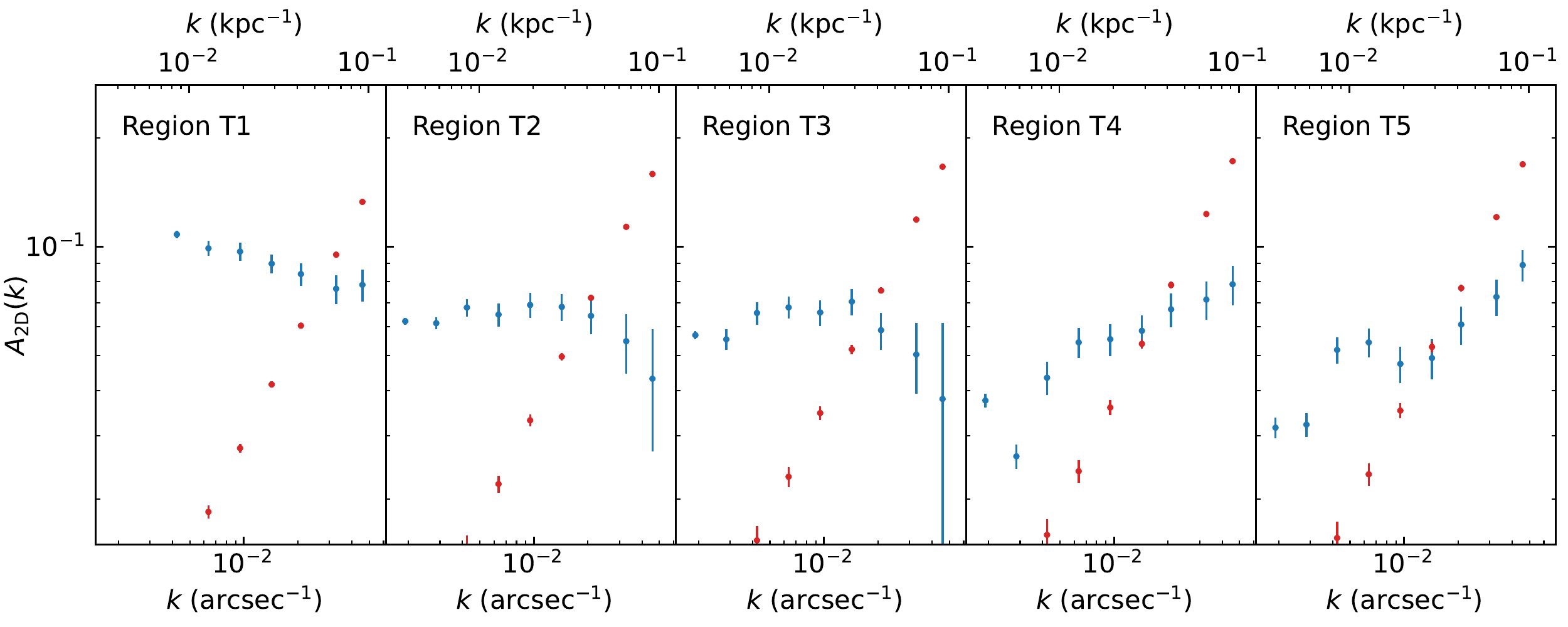}
    \vspace{-5mm}
    \caption{Top: 2D power spectrum of the surface brightness fluctuations (blue data points) for the five regions in the tail of the NGC 4839 group, with Poisson noise being subtracted. In each panel, the red data points are the subtracted Poisson noise, and the solid blue line is the best-fitting power-law model to the data, with covariance taken into account. The best-fitting values of the slope $\alpha$ with the 68 per cent confidence interval are shown in the upper-right corner. For all regions, the slope of the power spectrum is shallower than that of the Kolmogorov 2D power spectrum (green solid line), indicating suppression of thermal conduction throughout the tail. Bottom: Characteristic 2D amplitude of the surface brightness fluctuations (blue data points). The red data points are Poisson noise. The observed amplitude is in the approximate range of 0.03--0.10, depending on the scale, suggesting a mild level of turbulence throughout the tail.}  
    \label{fig:Tail_PS}
\end{figure*}

\begin{equation}
I_{\sigma} = \left ( \frac{G_{\sigma_{1}}\circ I }{G_{\sigma_{1}}\circ E } - \frac{G_{\sigma_{2}}\circ I }{G_{\sigma_{2}}\circ E } \right ) \times E ,
\label{equ: I_sigma}
\end{equation}

where $G_{\sigma_{1}}$ and $G_{\sigma_{2}}$ represent, respectively, Gaussian filters of widths $\sigma_1=\sigma/\sqrt{1+\varepsilon}$ and $\sigma_2=\sigma\sqrt{1+\varepsilon}$, with $\varepsilon \ll 1$. Following \cite{Arevalo2012}, we fixed $\varepsilon$  at $10^{-3}$. The variance of the convolved image ($I_\sigma$) for a given wave number $k$ is $V_k=\sum I^2_\sigma$, and is linearly proportional to the power spectrum \citep{Churazov2012}.



We convolved the count image by the modified filter (equation \ref{equ: I_sigma}) on varying scales from 5--120 pixels. Following \cite{Eckert2017}, we binned the \textit{Chandra} data by a factor of 2 so that 1 bin $= 0.984$ arcsec. We calculated the mean surface brightness, $\mu_S$, in each region so that the scaled 2D power spectrum can be given by the following \citep{Eckert2017}:
\begin{equation}
P_{\rm{2D}}(k)= \frac{1 }{\varepsilon^2 \pi k^2\mu^2_S}\frac{V_k}{\sum E^2},
\label{equ: P_2D}
\end{equation}
where the characteristic 2D amplitude of the fluctuations is defined as $A_{\rm{2D}}= \sqrt{P_{\rm{2D}}(k)2\pi k^2}$.

\subsubsection{Poisson noise and uncertainties}
\label{sec: Poisson}
The Poisson noise contribution is expected to be given by a flat power spectrum. This power can be calculated from the raw image as following \citep{Eckert2017}:
\begin{equation}
P_{\rm{Poisson}}(k)= \frac{1 }{\mu^2_S}\frac{\sum I}{\sum E^2}.
\label{equ: P_Poisson}
\end{equation}

As discussed in \citet{Churazov2012} and \citet{Eckert2017}, there is a correlation among the powers calculated at different scales, which introduces a covariance between data points that needs to be taken into account. In order to estimate the uncertainties and create the covariance matrix, we calculated the variance in each of the given regions. We then generated $N_{\rm{r}}=10,000$ realizations of the power spectrum by randomly shuffling the pixel values within each region. The covariance matrix was calculated by
\begin{equation}
\Sigma ^2_{i,j} = \frac{1}{N_{\rm{r}}}\sum ^{N_{\rm{r}}}_{\ell=1}(P_\ell(k_i)-\bar{P}(k_i))(P_\ell(k_j)-\bar{P}(k_j)),
\label{equ: covariance}
\end{equation}
where $P_\ell(k_i)$ and $P_\ell(k_j)$ represent the $\ell$th realization of the power spectra at scales $k_i$ and $k_j$, respectively. $\bar{P}(k_i)$ and $\bar{P}(k_j)$ are the mean power spectra. 


The 2D power spectrum and characteristic amplitude of the surface brightness fluctuations for each selected region in the tail of the NGC 4839 group are shown in Fig.~\ref{fig:Tail_PS}, with Poisson noise being subtracted. The uncertainties are taken from the square root of the diagonal elements of the covariance matrix. The figure shows that the power at smaller wave numbers (i.e. larger scale distances) decreases with increasing distance from the head of the group, while the power at higher wave numbers remains relatively the same. On the other hand, the characteristic amplitude of surface brightness fluctuations (Fig.~\ref{fig:Tail_PS}, bottom) peaks at a small wave number of $\sim 4 \times 10^{-3}$ arcsec$^{-1}$ in the inner tail (region T1), and as we move to the furthest part of the outer tail (such as regions T4 and T5), the amplitude peaks at a high wave number of $\sim 6 \times 10^{-2}$ arcsec$^{-1}$.





\subsubsection{Slope of the power spectrum}
As shown in Fig.~\ref{fig:Tail_PS} and by \cite{Eckert2017}, the shape of the power spectrum closely follows a power law. We, therefore, fitted the power spectrum for each selected region in the tail with a simple power law, $F(k)=P_0 k^{\alpha}$, where $P_0$ and $\alpha$ are the normalisation and slope, respectively. We used a likelihood function that takes the covariance between data points into account,
\begin{equation}
\textup{log} \mathscr{L} =- \frac{1}{2}\sum ^{N}_{i,j=1}(P_i-F(k_i))(P_j-F(k_j))(\Sigma ^2)^{-1}_{i,j},
\label{equ: likelihood}
\end{equation}
where $(\Sigma ^2)^{-1}_{i,j}$ is the inverse of the covariance matrix. For the fitting process, we used the affine-invariant ensemble sampler for Markov chain Monte Carlo implemented in the \texttt{EMCEE} package \citep{Foreman2013}. We chose a burn-in phase of 1,000 steps, followed by 10,000 MCMC steps.

The best-fitting power-law models to the extracted power spectra are shown as blue solid lines in the upper panel of Fig. \ref{fig:Tail_PS}. The best-fitting parameters and their 68 per cent uncertainties are presented in Table~\ref{tab: pow_law parameters}. Overall, the slope gets less steep as the distance from the head of the group increases. Near the head (region T1), it is the steepest with a slope of $\alpha=-2.35^{+0.07}_{-0.06}$ and steadily flattens with increasing distance to a value of $\alpha=-1.37^{+0.09}_{-0.07}$ at the furthest part of the tail (region T5). For all regions, our estimated slope of the power spectrum is shallower than the slope of the Kolmogorov 2D power spectrum ($\propto k^{-8/3}$).


\begin{table}
    \centering
     \caption{Best-fitting parameters of the power-law fit.}
        \begin{tabular}{c c|c|c}
     \hline
     Region & $\alpha$ & $P_0$ & Reduced $\chi^2$ \\
      \hline
      T1     & $-2.35_{-0.06}^{+0.07}$ &  $4.68_{-0.12}^{+0.14} \times 10^{-3}$ & 1.01 \\
      T2     & $-2.16_{-0.08}^{+0.10}$ &  $3.98_{-0.11}^{+0.11} \times 10^{-3}$ & 1.03 \\
      T3     & $-1.98_{-0.09}^{+0.10}$ &  $1.07_{-0.09}^{+0.08} \times 10^{-2}$ & 1.10 \\
      T4     & $-1.67_{-0.04}^{+0.04}$ &  $3.47_{-0.16}^{+0.16} \times 10^{-2}$ & 1.12 \\
      T5     & $-1.37_{-0.07}^{+0.09}$ &  $1.05_{-0.11}^{+0.10} \times 10^{-1}$ & 1.09 \\ 
        \hline
      \end{tabular}
     \label{tab: pow_law parameters}
\end{table}





\section{Discussion}
\label{sec: discussion}
\subsection{Transport processes}
\label{sec: transport processes}
Based on our findings presented in Section \ref{sec: tail}, the slope of the power spectrum for the surface brightness fluctuations in the tail of the NGC 4839 group gets less steep as the distance from the head of the group increases. The power at smaller wave numbers decreases as we move further away from the head, while the power at higher wave numbers stays relatively the same. This implies that the power from larger turbulent eddies decreases with increasing distance from the head, where the bulk of the turbulence is being generated as the group falls into the Coma cluster. In keeping with the energy cascade of turbulent motion (i.e. large eddies feeding small eddies), we should also see a decrease in contribution from larger turbulent eddies and an increase from smaller turbulent eddies as the distance from the head increases. The characteristic amplitude of surface brightness fluctuations (Fig.~\ref{fig:Tail_PS}, bottom) does indeed show this feature. In region T1, the amplitude peaks at a small wave number, and as one moves to the outermost region, region T5, the amplitude peaks at a high wave number.

\begin{figure}
    \centering
    \includegraphics[width=\columnwidth]{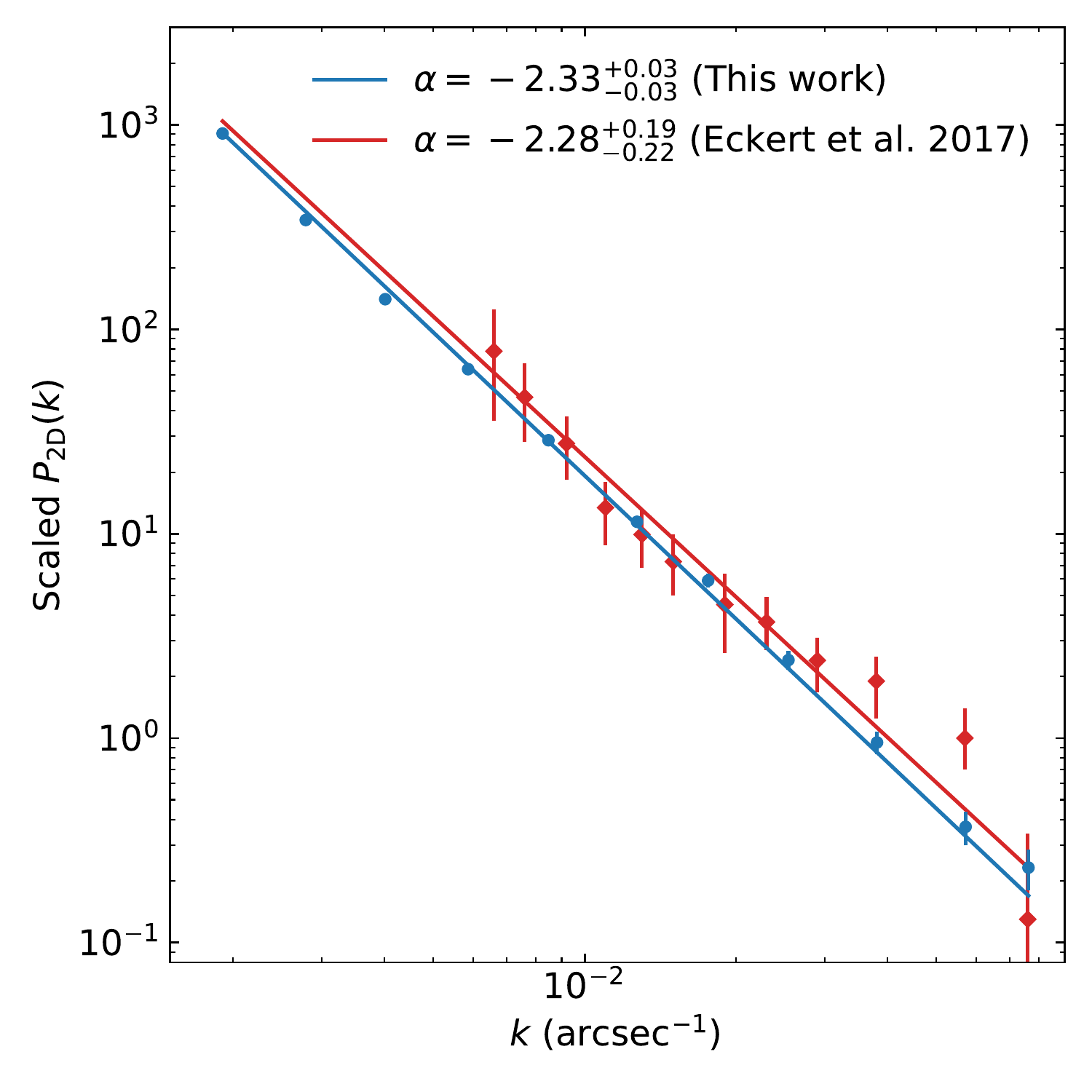}
    \vspace{-7mm}
    \caption{Scaled 2D power spectrum of the surface brightness fluctuations for the entire tail of the NGC 4839 group, compared to that of the infalling group in Abell 2142 \citep{Eckert2017}. Our estimated slope, which is in agreement with that estimated for the infalling group in Abell 2142, is flatter than the Kolmogorov slope, indicating that thermal conduction is being suppressed throughout the tail.}
    \vspace{-4mm}
    \label{fig:Whole_power}
\end{figure}

Using \textit{Chandra} observations, \citet{Eckert2017} determined the power spectrum of surface brightness fluctuations in the whole tail of the infalling group in Abell 2142 to reveal the microphysics of its gas. In order to have a proper comparison to the power spectrum of the infalling group in Abell 2142, we constructed a power spectrum for the entire tail of the NGC 4839 group (Fig.~\ref{fig:Whole_power}). We found a slope of $\alpha=-2.33 \pm 0.03$, which is in agreement with the slope of $\alpha=-2.28^{+0.19}_{-0.22}$ for the tail of the infalling group in Abell 2142. In both cases, the slope is flatter than the Kolmogorov slope. \cite{Eckert2017} attributed this to the suppression of conductivity within the bulk of the plasma (\citealt{Gaspari13}). Such suppression results in the slow mixing of the merging plasma and the long survival of small-scale fluctuations. The NGC 4839 group exhibits both of these qualities: the temperature within the tail stays above that of the surrounding ICM for several hundred kiloparsecs (see Fig.~\ref{fig:temp_map}) and the power of small-scale fluctuations is relatively constant throughout the tail. If thermal conduction was not inhibited, Spitzer-like thermal conduction would wash out small-scale perturbations in the ICM, resulting in steeper power spectrum (\citealt{Gaspari14}). Therefore, like the infalling group in Abell 2142, our findings suggest that thermal conduction is suppressed throughout the tail of the NGC 4839 group.

As can be seen in our \textit{Chandra} image of the group (Fig. \ref{fig:flux_image}, right-hand panel), the morphology of the tail is complex, but overall can be described by a 3D conical geometry. For a conical geometry and an inclination angle of 30 degree or smaller with respect to the plane of the sky, \citet{Eckert2017} have shown that the normalization of the 3D power spectrum, $P_{\rm{3D}}(k)$, changes by a factor of 4 when converting from the projected 2D power spectrum, $P_{\rm{2D}}(k)$. As shown in the bottom panel of Fig. \ref{fig:Tail_PS}, the characteristic 2D amplitude, $A_{\rm{2D}}(k)$, is in the approximate range of 0.03--0.10 throughout the entire range of wave numbers. Using these values and the conversion factor between $P_{\rm{2D}}(k)$ and $P_{\rm{3D}}(k)$, we found that the characteristic 3D amplitude, $A_{\rm{3D}}=\sqrt{P_{\rm{3D}}(k) 4 \pi k^{3}}$, is in the range of 0.02--0.07, suggesting a mild level of turbulence with a Mach number in the range of 0.1--0.5 \citep{Gaspari13}.

\subsection{Group dynamics}
\label{sec: dynamics}
Cosmological simulations \citep{Lyskova2019,Sheardown19,Zhang19b} showed that the NGC 4839 group initially passed through the Coma centre from the northeast and reversed its direction of motion after reaching the apocentre, and is now heading back for its second infall. Recent observations of the Coma cluster with \textit{eROSITA} \citep{Churazov2021} and \textit{LOFAR} \citep{Bonafede21} reveal evidence of a faint bridge connecting the NGC 4839 group with the Coma cluster, indicating that the group has already passed the main cluster. The presence of a shock \citep[e.g.][]{Simi13,Akamatsu2013,Churazov2022arXiv} at the outer edge of the radio relic to the southwest of the group \citep[e.g.][]{Brown2011,Bonafede2022} further supports the post-merger scenario. In the left-hand panel of Fig.~\ref{fig:flux_image}, we show the locations of this shock (labelled as primary shock) and the radio relic.  

Previous X-ray studies \citep[e.g.][]{Neumann2003,Simi13,Mira20} identified a sharp edge to the west of the Coma core, extending to the region connecting the group to the Coma cluster (see Fig.~\ref{fig:flux_image}, left). Using X-ray data from \textit{eROSITA}, \citet{Churazov2021} interpreted this edge as a "secondary" or "mini-accretion" shock driven by the first passage of the NGC 4839 group through the Coma core. Based on the numerical simulations of \citet{Zhang2021Pairs}, \citet{Churazov2021} argued that this "secondary" shock and the one at the outer edge of the radio relic are caused by the NGC 4839 group. According to this scenario, the gas displaced by the first passage of the NGC 4839 group eventually falls back and reaches hydrostatic equilibrium, forming a "secondary" shock. While the primary merger shock continues to propagate outwards and is currently located at the position of the radio relic. 


Our \textit{Chandra} observations of the group reveal the presence of cold and shock fronts at the interface connecting the group to the Coma cluster, indicating that the group is heading radially towards the Coma centre. Based on our spatial and spectral analyses of the leading head, we can determine the infall velocity of the group from the bow shock. Using an ambient gas temperature of 3 keV, the local speed of sound is $c_s=1480(T/10^8 {\rm{K}})^{1/2} = 875$ km s$^{-1}$. For a Mach number of $1.5_{-0.5}^{+0.8}$ estimated using the gas temperature, the infall velocity of the group is $\sigma_v = c_s  \mathcal{M}= 1315_{-440}^{+700}$ km s$^{-1}$ relative to the Coma cluster. The estimated infall velocity drops to $1230_{-180}^{+265}$ km s$^{-1}$ if we adopt a Mach number of $1.4_{-0.2}^{+0.3}$ estimated using the gas density. These values are consistent with the infall velocity of $1700_{-500}^{+350}$ km s$^{-1}$ estimated for the NGC 4839 group based on galaxy redshifts and a dynamical two-body model \citep{CD96}.

\section{Conclusions}
\label{sec: conclusions}
Based on our new deep \textit{Chandra} observations, we have carried out a detailed analysis of the NGC 4839 group falling into the Coma cluster. The \textit{Chandra} observations of the infalling group reveal a very complex morphology with a long tail of ram-pressure stripped gas extending for several hundred kiloparsecs, preceded by an X-ray bright head-like structure. The main findings of this work can be summarised as follows:

\begin{enumerate}
 \item Our analysis reveals a cold front feature with a very high density jump at the leading head of the group, preceded by a bow shock of hot gas to the southeast following the ridge of emission. At this shock front, the gas temperature drops by a factor of $1.7$, corresponding to a Mach number of $1.5_{-0.5}^{+0.8}$, statistically consistent with a Mach number of $1.4_{-0.2}^{+0.3}$ estimated using the gas density. 
 \item The power spectrum of surface brightness fluctuations in the tail shows that the power at larger scale distances, i.e. smaller wave numbers, decreases with increasing distance from the head of the group, while the power at smaller scale distances remains relatively the same. This implies that the power from larger turbulent eddies decreases with distance from the head, where the bulk of the turbulence is generated as the group falls into the main cluster.
 \item The slope of the power spectrum of surface brightness fluctuations in the tail of the NGC 4839 group gets less steep as the distance from the head increases, changing from $\alpha=-2.35_{-0.06}^{+0.07}$ at the inner part of the tail to $\alpha=-1.37_{-0.07}^{+0.09}$ at the outermost part of the tail. The power spectrum for the entire tail has a slope of $\alpha=-2.33^{+0.03}_{-0.03}$, consistent with the slope found in the tail of the infalling group in Abell 2142 \citep[$\alpha=-2.28^{+0.19}_{-0.22}$;][]{Eckert2017}. Our estimated slopes are shallower than the slope of the Kolmogorov 2D power spectrum, indicating that thermal conduction is being suppressed throughout the tail.
 \item The characteristic 2D amplitude of surface brightness fluctuations in the tail peaks at large-scale distances in the inner tail, while as we move to the outermost part of the tail, the amplitude peaks at small-scale distances. This implies an increase in contribution from smaller turbulent eddies and a decrease from larger eddies as the distance from the head increases. We found that the characteristic 2D amplitude is in the approximate range of 0.03--0.10 throughout the entire range of wave numbers, suggesting a mild level of turbulence with a Mach number in the range of 0.1--0.5 \citep{Gaspari13}.
 \item Our \textit{Chandra} observations of the group reveal a KHI-like feature along the northern side of the inner tail. The GGM-filtered images of the group show the gradient in surface brightness along this side is not as sharp as that along the southern side of the inner tail, which could indicate the presence of KHI rolls that mix the stripped gas with the ambient ICM. On the other hand, the X-ray bright tail that extends for several hundred kiloparsecs in the wake argues against the existence of KHI rolls, which might indicate significant viscosity or draped magnetic fields. As an alternative, this peculiar structure could be due to the complex dynamic history of the group. However, given the small spatial extent of this structure, the existing data do not allow us to conclusively determine its nature.  

\end{enumerate}

\section*{Acknowledgements}
We thank the referee for their report. MSM, SAW and JR acknowledge support from \textit{Chandra} grant GO0-21120X. This work is based on observations obtained
with the \textit{Chandra} observatory, a NASA mission.

\section*{Data Availability}
The \textit{Chandra} Data Archive stores the data used in this paper. The \textit{Chandra} data were processed using the \textit{Chandra} Interactive Analysis of Observations (\texttt{CIAO}) software. The \textit{XMM--Newton} Science Archive (XSA) stores the archival data used in this paper, from which the data are publicly available for download. The \textit{XMM} data were processed using the \textit{XMM--Newton} Science Analysis System (\texttt{SAS}). The software packages \texttt{HEASoft} and \texttt{XSPEC} were used, and these can be downloaded from the High Energy Astrophysics Science Archive Research Centre (HEASARC) software web page.



\bibliographystyle{mnras}
\bibliography{NGC4839} 


\bsp	
\label{lastpage}
\end{document}